\documentclass[aps,prl,twocolumn,showpacs,superscriptaddress,groupedaddress]{revtex4}  
\usepackage{graphicx}  
\usepackage{dcolumn}   
\usepackage{bm}        
\usepackage{amssymb}   

\begin{document}



\title{Oriented rotational wave-packet dynamics studies via high harmonic generation.}
\author{E. Frumker}
\email{eugene.frumker@mpq.mpg.de}
\affiliation{Joint Attosecond Science Laboratory, University of Ottawa and National Research Council of Canada, 100 Sussex Drive, Ottawa, ON K1A 0R6, Canada}
\affiliation{Max-Planck-Institut f\"{u}r Quantenoptik, Hans-Kopfermann-Strasse 1, D-85748 Garching, Germany}
\affiliation{Department of Physics, Texas A\&M University, College Station,Texas 77843, USA}

\author{C. T. Hebeisen}
\affiliation{Joint Attosecond Science Laboratory, University of Ottawa and National Research Council of Canada, 100 Sussex Drive, Ottawa, ON K1A 0R6, Canada}
\author{N. Kajumba}
\affiliation{Joint Attosecond Science Laboratory, University of Ottawa and National Research Council of Canada, 100 Sussex Drive, Ottawa, ON K1A 0R6, Canada}
\affiliation{Department f\"{u}r Physik der Ludwig-Maximilians-Universit\"{a}t, Schellingstrasse 4, D-80799 Munich, Germany}
\author{J. B. Bertrand}
\affiliation{Joint Attosecond Science Laboratory, University of Ottawa and National Research Council of Canada, 100 Sussex Drive, Ottawa, ON K1A 0R6, Canada}
\author{H. J. W\"{o}rner}
\affiliation{Joint Attosecond Science Laboratory, University of Ottawa and National Research Council of Canada, 100 Sussex Drive, Ottawa, ON K1A 0R6, Canada}
\affiliation{Laboratorium f\"{u}r physikalische Chemie, ETH Z\"{u}rich, Wolfgang-Pauli-Strasse 10, 8093 Z\"{u}rich, Switzerland}
\author{M. Spanner}
\affiliation{Steacie Institute for Molecular Sciences, National Research Council of Canada, Ottawa, ON  K1A 0R6, Canada}
\author{D. M. Villeneuve}
\affiliation{Joint Attosecond Science Laboratory, University of Ottawa and National Research Council of Canada, 100 Sussex Drive, Ottawa, ON K1A 0R6, Canada}
\author{A. Naumov}
\affiliation{Joint Attosecond Science Laboratory, University of Ottawa and National Research Council of Canada, 100 Sussex Drive, Ottawa, ON K1A 0R6, Canada}
\author{P.B. Corkum}
\email{paul.corkum@nrc.ca}
\affiliation{Joint Attosecond Science Laboratory, University of Ottawa and National Research Council of Canada, 100 Sussex Drive, Ottawa, ON K1A 0R6, Canada}


\begin{abstract}
We produce oriented rotational wave packets in CO and measure their characteristics via high harmonic generation. The wavepacket is created using an intense, femtosecond laser pulse and its second harmonic. A delayed 800 nm  pulse probes the wave packet, generating even-order high harmonics that arise from the broken symmetry induced by the orientation dynamics.  The even-order harmonic radiation that we measure appears on a zero background, enabling us to accurately follow the temporal evolution of the wave packet. Our measurements reveal that, for the conditions optimum for harmonic generation, the orientation is produced by preferential ionization which depletes the sample of molecules of one orientation.
\end{abstract}

\maketitle




Laser-assisted molecular alignment has become an enabling tool for molecular frame studies in physics and chemistry \cite{Stapelfeldt_Aligning_molecules_RevModPhys_2011}. Field-free molecular alignment plays a particularly important role in attosecond and recollision science, facilitating, for example, high harmonic orbital tomography \cite{Itatani_tomog_Nature2004,Vozzi_gen_tomography_NatPhys_2011}, and laser induced electron diffraction experiments \cite{Meckel_laser_diffraction_Science2008}. For polar molecules, in addition to alignment, it is important to orient the sample to avoid averaging over two opposite molecular orientations.

Since the work of Friedrich and Herschach \cite{Friedrich_Orientation_NATURE_2001}  on orientation of molecules in the presence of a strong DC field, several field-free orientation techniques have been demonstrated. Orientation methods include using a combination of a static electric field and a non-resonant femtosecond laser excitation \cite{Sakai_control_orient_PRL_2003,Holmegaard_Orient_large_Mol_PRL_2009,  Ghafur_Orientation_Nat_Phys_2009}; using a combination of an electrostatic field and an intense nonresonant rapidly turned-off laser field \cite{Goban_Sakai_Orientation_PRL_2008}; using an IR and UV pulse pair \cite{Sokolov_Scully_orientation_PRA_2009}; using THz pulses \cite{Fleischer_Orient_THz_PRL2011}; and using two-color laser fields \cite{Vrakking_orientation_2colour_CPL_1997, De_Orientation_w2w_PRL2009}.   To date orientation has been achieved in low gas densities \cite{Holmegaard_Orient_large_Mol_PRL_2009} or with such a small degree of orientation \cite{Sakai_control_orient_PRL_2003, Holmegaard_Orient_large_Mol_PRL_2009, Fleischer_Orient_THz_PRL2011} that high harmonics experiments have not been feasible.

We demonstrate field-free orientation of CO molecules by using two-color ($\omega+2\omega$) laser fields. The degree of orientation of the sample is probed by a second, delayed, probe pulse that generates high harmonic radiation (HHG).  HHG in an isotropic gas produces odd harmonics of the laser frequency because alternate half-cycles of the laser field produce attosecond pulses with alternating sign.  Anisotropy in the medium, produced for example by oriented molecules, breaks the strict alternation of sign, leading to the appearance of even harmonics of the laser frequency. Because there is no background at the spectral position of the even harmonics, HHG is a very sensitive measure of orientation.

There are two contributing mechanisms of two-color laser fields that lead to orientation in polar molecules.  One is that of the hyperpolarizability interaction \cite{Kanai_w2w_simulations_JCP_2001,De_Orientation_w2w_PRL2009}. The other is depletion, in which the two-color field preferentially ionizes the part of the isotropic distribution that is oriented in a particular direction relative to the laser polarization, leading to a hole in the distribution of neutral molecules.  In both cases, the laser gives a kick to the molecules, leading to rotational revivals as is commonly used to align molecules \cite{Stapelfeldt_Aligning_molecules_RevModPhys_2011}. We will show that at the condition optimum for harmonic generation, the depletion mechanism dominates leading to substantial asymmetry.

\begin{figure}[t]
	\centering
	\includegraphics[width=\columnwidth]{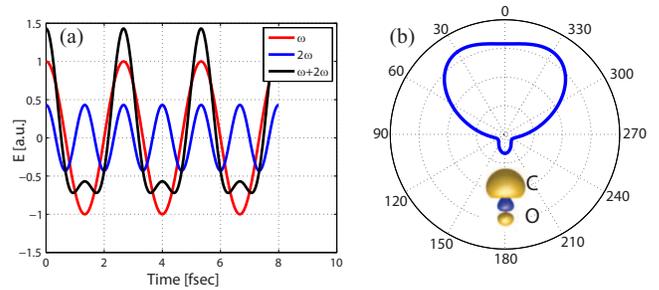}
	\caption{\textbf{(a)} Plot of two-color laser fields: Red line $E_{\omega}\cos(\omega t)$ - fundamental field. Blue line $E_{2\omega}\cos(2\omega t+\varphi_{2\omega})$ - second harmonic. Black line $E_{total}(t)=E_{\omega}\cos(\omega t)+E_{2\omega}\cos(2\omega t+\varphi_{2\omega})$, $\varphi_{2\omega}=0$, $E_{2\omega}=0.43E_{\omega}$  - the total field in the pump beam. The relative field ratio shown in the plot is the same as used in our experiment. \textbf{(b)} Half cycle total ionization rate of CO molecule as a function of angle between the molecular axis and the electric force calculated for intensity $1.5\times10^{14}$ $\textrm{W}/\textrm{cm}^2$ at $800$ nm.  -C- side points toward $0^{\circ}$. }
	\label{Fig1_w2w}
\end{figure}

Figure \ref{Fig1_w2w} illustrates the concept of orientation by ionization.  A superposition of an $\omega$ and a $2\omega$ pulse are focused into a thin gas jet (thin relative to the Raleigh range) of either beam.  When their relative phase $\varphi_{2\omega}=0, \pi, 2\pi,$ etc. the resulting electric field of the pump is maximally asymmetric as sketched in the figure.  The tunnel ionization probability can be highly sensitive to molecular orientation in polar molecules \cite{Akagi_tunnel_ion_Science2009,Stapelfeldt_OCS_orient_NatPhys2010}. In the polar plot, we show the angle dependence of the ionization rate for a $1.5\times10^{14}$ $\textrm{W} / \textrm{cm}^2$ pulse as a function of molecular frame angle.  The ionization probability is 5 times greater when the electrical force points to the C side of the molecule than when it points to the O side.  Thus, molecules with the C atom pointing opposite to the direction of the field maximum of the pump pulse will ionize more readily than any other.  This depletes the sample of one orientation of molecules.  The same pulse that depletes the sample creates a rotational wave packet.  Thus, we have created a preferentially oriented neutral rotational wave packet and potentially an oriented rotational wave packet in the ion.

Using high harmonic generation as a diagnostic makes us largely blind to the ionized molecules.  They have a much higher ionization potential ($I_p^{ion}$=26.8 eV) than the neutrals ($I_p^{ntrl}$=14 eV).  In addition, ionization creates vibrationally excited ions that will lead to rapid dephasing of the rotational wave packet.  After only one revival we cannot observe a well-defined ion wave packet in the ion in companion Coulomb explosion experiments \cite{Heibesen_CEI_2012}.  Effectively we can think that preferential ionization burns an asymmetric hole in otherwise symmetric distribution of molecules.

\begin{figure}[t]
	\centering
\includegraphics[width=\columnwidth]{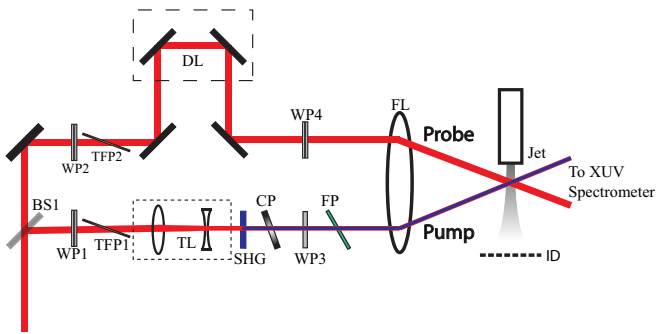}
	\caption{Schematic of experimental setup. A $\lambda/2$ waveplate (WP1,WP2) and a thin film polarizer (TFP1, TFP2) were used in the pump and the probe arms to continuously control the pulse energies. The pump beam was de-magnified (1.5:1 telescope (TL)) to ensure full overlap with the probe beam when focused on the jet.
The probe beam was delayed using a computer-controlled delay line (DL). The parallel, but spatially separated, pump and probe beams impinge on a spherical $\textrm{f}=500 $ mm focusing mirror (FL, shown as a lens for clarity). The pump and probe beams overlap on the gas jet with a small angle (1$^\circ$) between  their propagation directions.}
	\label{Fig_setup}
\end{figure}

The schematic of our experimental set-up is shown in Fig. \ref{Fig_setup}.  A linearly polarized laser pulse from a Titanium:Sapphire laser ($\sim$35 fs,
$\lambda$=800 nm, 50 Hz, $\sim$10 mJ)  was split by beamsplitter (BS1) into pump and probe arms.
In the probe arm, the second harmonic beam ($\lambda$=400 nm) was generated in a 200 $\mu$m Type-I BBO crystal (SHG).
 By means of a birefringent calcite plate (CP) and a zero order half-wave plate (WP3), the 800
and 400 nm pulses are made coincident in time and parallel in polarization.
 The phase between the $\omega$ and $2\omega$ fields is controlled by a 1 mm fused silica plate (FP).


The sample of rotationally cold CO molecules is produced by a supersonic expansion
from a pulsed valve composed of $10\%$ CO  diluted in helium.  The gas jet thickness ($0.5$ mm) is considerably smaller than
the confocal parameter ($20$ mm) of the pump beam, removing any effect of the Gouy phase shift.

  Both the pump and probe harmonics were analyzed in an XUV flat field spectrometer.  The details of this spectrometer are described elsewhere \cite{Frumker_sword_OL2009}, and detected by an imaging multi-channel plate (MCP) / cooled charge-coupled device (CCD) camera combination.
  To ensure the greatest possible linearity of the detector and to facilitate extended dynamic range data acquisition, the MCP detector was operated at a low voltage (1700 V across the MCP chevron pair) and the same image was acquired by the cooled CCD multiple times with different integration times. We determined the linear range of the camera with respect to the integrated intensity and masked out saturated and non-linear regions of each image. The remaining parts of the images were integrated and calibrated according to the total acquisition time of each pixel. This method provides substantial noise reduction and an extended dynamic range over single acquisitions.

\begin{figure}[t]
	\centering
	\includegraphics[width=\columnwidth]{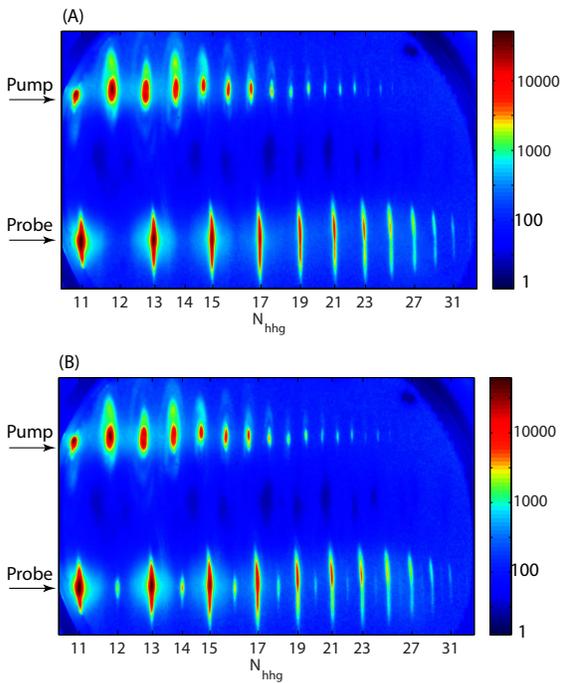}
	\caption{Logarithmic image of the measured high harmonic signal (along the line marked as ``Probe'') and harmonics produced by the pump beam (along the line marked as ``Pump''). (A) No orientation $t_{\textrm{delay}}=7200$ fsec.  (B)  Maximum orientation $t_{\textrm{delay}}=8850$ fsec.}
	\label{Fig_full_image}
\end{figure}

The non-collinear beam geometry enables us to spatially separate harmonics produced by the alignment/
orientation (pump) beam from those produced by the probe beam.  Figure \ref{Fig_full_image} shows the harmonics generated by the pump beam and the probe beam.  The pump beam is off axis and its harmonics are labelled (Pump), while the harmonics generated in the probe beam are labelled (Probe)  in the figure.  Since the off-axis high harmonics are produced by the pump, we know that the medium is being ionized. (Even harmonics produced through a two color field are well known \cite{Watanabe_w2w_control_PRL_1994}.)  Fig. \ref{Fig_full_image} (a) and (b)  shows CCD images taken at the pump-probe delay of 7200 fsec ( no orientation) and 8850 fsec (maximum orientation) respectively. The whole new set of even harmonics in the probe beam appears when the medium is oriented (Fig. \ref{Fig_full_image} (b))

From the cut-off ($n_{\textrm{cutoff}}=31$) of the signal harmonics we determine the peak intensity in the probe beam to be $\sim 1.8\times10^{14}$ $\textrm{W}/\textrm{cm}^2$. We measure the relative beam energies of the pump ($\omega$ and $2\omega$ components separately) and probe beam just before the focusing mirror. Since we can estimate the beam sizes and the pulse durations at the jet, we know the pump beam peak intensities of $\omega$ and $2\omega$ at the jet. We observed best even harmonic signal from the probe for pump intensities $I_{\omega}\simeq1.46\times10^{14}$ $\textrm{W}/\textrm{cm}^2$ and $I_{2\omega}\simeq2.73\times10^{13}$ $\textrm{W}/\textrm{cm}^2$.

We also observe the ion signal produced by the pump and the probe beams in the interaction region, using an ion detector (shown as (ID) in Fig. \ref{Fig_setup}) . We measure that the ion signal in the pump ($\omega+2\omega$) arm is about 20 times larger than the ions signal produced by the probe beam ($\omega$ only), just when high harmonics become experimentally observable.
Taking into account the focal volumes of pump and probe beams, we estimate that the generating  medium in our experiments was ionized by the pump beam on the order of $30\%$.

There are two reasons why ionization could break the symmetry and permit the probe pulse to create even harmonics. It could create an oriented ensemble, as we described in our introduction. However it will also induce a plasma current, which will result in a charge separation and a plasma field that also breaks symmetry.
To experimentally confirm that the even harmonics were caused by orientation of the gas sample, we replaced $\textrm{CO}$ with $\textrm{N}_2$ at the otherwise identical experimental conditions. The even harmonics disappeared ruling out this plasma effect.

Now, we turn our attention to the harmonics in Fig. \ref{Fig_full_image} created by the probe beam. The most prominent feature is the appearance of even harmonics at $t\sim 8850$ $\textrm{fsec}$ after the pump pulse. This is very close to the full revival time of a $\textrm{CO}$ rotational wave packet. We will show below that the detailed wave packet dynamics near the full revival will help to further clarify the mechanism  of molecular orientation.

Lowering the pump pulse intensity by attenuating the fundamental component of the pump beam by about $10\%$ made even harmonics in the probe arm in Fig. \ref{Fig_full_image} disappear, while having little effect on the odd harmonics in the spectrum. (Since the second harmonic conversion was not saturated we estimate that the second harmonic of the pump is attenuated by $20\%$.) In contrast, when lowering the probe pulse intensity, the even harmonics created by the pump didn't disappear from the spectrum, while the cut-off and the overall intensity of the probe harmonics was reduced.
Further increasing pump intensity above the optimum is also deleterious.

The odd harmonics created by the probe provide a reference against which we can evaluate the even harmonics.  In Fig. \ref{Fig_full_image} (b) the even harmonic signal is approximately $5\%$ of the odd harmonic signal.


  Since high harmonic generation is a coherent process, the harmonic signal is proportional to the square of the number of emitters.
  Therefore, to a first approximation, the ratio between even and odd harmonic intensity ($\textrm{I}_{even}/\textrm{I}_{odd}$) is proportional to the square of the degree of orientation ($\textrm{I}_{even}/\textrm{I}_{odd}\sim\langle \cos\theta\rangle^2$).
 Averaged over the spectrum, we measure $\textrm{I}_{even}/\textrm{I}_{odd}\simeq0.05$. This ratio roughly places the degree of orientation $\langle \cos\theta\rangle\simeq0.20$. This means that the harmonic strength is as if we had $20\%$ of molecules perfectly oriented (or more partially oriented).

%
\begin{figure}[t]
	\centering
	\includegraphics[width=\columnwidth]{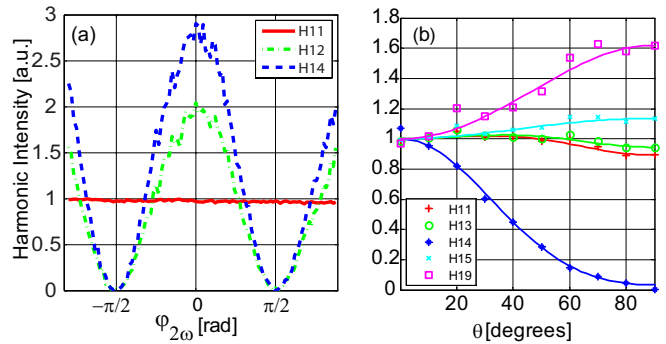}
	\caption{:  (a) Measured even and odd harmonic signal as a function of the phase delay $\varphi_{2\omega}$ between fundamental and second harmonic fields. (b) Plot of the measured angle dependence of a selection of harmonics taken at the time of maximum orientation by changing the angle $\theta$ of the probe polarization relative to the orientation angle.    }
	\label{Fig_rotation_wavepacket}
\end{figure}

Figure \ref{Fig_rotation_wavepacket} (a) concentrates on $t=8850$ $\textrm{fsec}$, showing one of the normalized odd harmonics  as well as the even harmonic signal for selected harmonics as a function of the relative phase of the  $\omega$ and $2\omega$  fields measured in the units of the $\varphi_{2\omega}$. All even harmonics maximize at the same relative phase. We can continuously control  $\varphi_{2\omega}$  by rotating the fused silica plate shown as (FP) in Fig.2. When  $\varphi_{2\omega}=\pm\pi/2$  the two-color pump pulse becomes symmetric and no orientation is created in the sample. The even harmonics vanish but, as seen from the figure, there is practically no effect on the odd harmonics. This agrees with the theoretical prediction \cite{Frumker_ABI_2012}, showing that the emission of odd harmonics is independent of the degree of orientation, and depends only on the degree of alignment. We attribute the small monotonic reduction in the odd harmonic signal ($<5\%$ barely visible on the picture scale) to the angle change of the pump beam when rotating the (FS) and the resulting small change of the phase matching condition. A phase delay of  $\varphi_{2\omega}=2\pi$  inverts the field direction in the pump pulse creating an identical wave packet but oriented in the opposite direction.  This imposes a phase inversion to the harmonic signals, but our measurements are insensitive to such an inversion.

Figure \ref{Fig_rotation_wavepacket}(b) shows harmonics H14 and H15 observed at the time of maximum orientation measured as a function of the angle   between the laser polarization and the alignment axis.  In our experiment, control of  angle $\theta$, is achieved using the $\lambda/2$ waveplate (WP4) shown in Fig. \ref{Fig_setup}. To neutralize the influence of the polarization direction on the spectrometer efficiency we normalize the measured signal to the signal without the pump pulse. The signal for H15 increases with the angle.  We observe the same behaviour for all odd harmonics in CO except the lowest (H11 and H13). (This contrasts with $\textrm{N}_2$, a molecule isoelectronic with CO.  For $\textrm{N}_2$, except at the high harmonic cut-off, the harmonic signal peaks when the molecule aligns).

The general behavior of H14 is the same as that of all other even harmonic that we measure (H12-H30).  We see that the harmonic signal falls continuously from a peak when the molecule is parallel to the field direction of the fundamental beam and becomes unobservable when the molecule is perpendicular. This verifies that the orientation is indeed produced by the optical field, and not, for example, by charge separation due to the asymmetric ionization probability.

We now turn to wave packet dynamics.  A companion theoretical paper \cite{Spanner_Orient_Mech20122} shows two quite different revival structures for an orientational wave packet created by an $\omega+2\omega$ pulse interacting with polar molecules depending upon whether the wave packet is launched through molecular hyperpolarizability  or depletion via ionization.
 Without the action of alignment forces (the interaction of the laser electric field with the linear part of the induced dipole moment), depletion creates orientation immediately by ionizing away unwanted molecules. However, the pump pulse  will also align the molecules and the presence of  alignment acting simultaneously with depletion sharpens the instantaneously created orientation ensemble along the pump field polarization axis shortly after $t=0$. At the full revival, these two contributions result in maximum  orientation after the full revival time. In contrast, a molecule must rotate before the wave packet displays its orientation if the interaction occurs through hyperpolarizability. The interplay between alignment through the polarisability and orientation by hyperpolarizability causes the orientation to reach its maximum at times earlier than the full revival as shown in ref. \cite{Spanner_Orient_Mech20122}.

%

\begin{figure}[t]
	\centering
\includegraphics[width=\columnwidth]{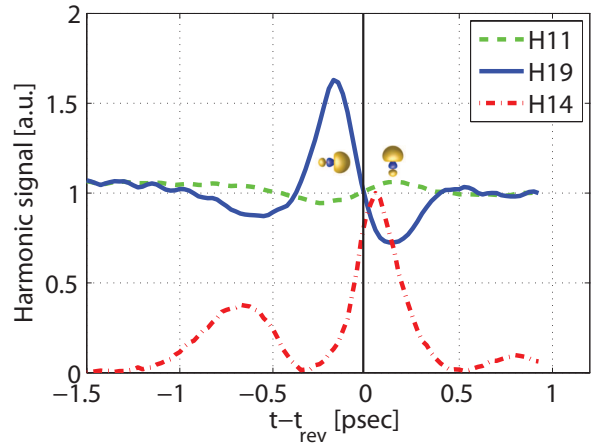} 

	\caption{Upper plot: A rotational alignment wave packet measured in CO excited by impulsive alignment with a superposition of 800 nm and 400 nm laser pulses.  Using this rotational wave packet the time dependence of the 11th harmonic (dashed green curve) and the 19th harmonic (solid blue curve) in $\mathrm{CO}$ is shown near the time of their first full revival $t_{rev}$ (set at $t_{rev}=0$). The harmonic signal normalized to the base line. Lower plot: An orientational wave packet in CO excited by the same superposition of 800 nm and 400 nm laser pulses. The red dash-dotted curve shows the time dependent signal for the 14th harmonic.}
	\label{Fig3_revival_experimental}
\end{figure}


Figure \ref{Fig3_revival_experimental} shows typical measured normalized signals for even (H14), and odd (H11, H19) harmonics plotted as a function of time. The H11 and H19 signals exhibit opposite revival structure, and their crossing at $t=0$ confirms that our measured revival time is correct.

The measured even harmonic signal reaches its maximum $(75\pm15)$ fsec after the revival of the alignment signal (Fig. \ref{Fig3_revival_experimental}). The delay and overall shape of the time evolution of the  orientation wavepacket signal agrees well with the preferential ionization mechanism.


In conclusion, we have shown that neutral, polar molecules such as CO can be oriented by depleting neutral gas of one molecular orientation using multiphoton ionization in an intense  $(\omega+2\omega)$  pulse.  High harmonic generation is an effective way to study oriented molecules.  This opens a route for applying high harmonic spectroscopy \cite{Woerner_Chem_Nature_2011} and orbital tomography \cite{Itatani_tomog_Nature2004} to polar molecules.

We acknowledge financial support of Canada's National
Research Council, National Sciences, Engineering
Research Council, the Canada Research Chair Program
and MURI grant W911NF-07-1-0475. E.F acknowledges
the support of Marie Curie International Outgoing Fellowship.


\bibliographystyle{unsrt}

\end{document}